\documentclass[showpacs,amsmath,amssymb,aps,prb,lengthcheck,superscriptaddress,twocolumn]{revtex4}
\usepackage[T1]{fontenc}
\usepackage[latin1]{inputenc}
\usepackage{amsmath}
\usepackage{graphicx}
\usepackage{epsfig}
\usepackage{bm}
\usepackage{dsfont}
\usepackage{color}
\usepackage{braket}

\newcommand{\nn}{\nonumber}

\newcommand{\bea}{\begin{eqnarray}}
\newcommand{\ea}{\end{eqnarray}}
\newcommand{\beq}{\begin{equation}}
\newcommand{\eq}{\end{equation}}
\newcommand{\bc}{\begin{center}}
\newcommand{\ec}{\end{center}}

\graphicspath{{./Figures/}} 

\begin{document}

\author{Damian Wozniak}
\affiliation{Department of Physics, Royal Holloway, University of London,
 Egham, Surrey TW20 0EX, United Kingdom}
\title{Manipulating the Mott lobes: optical lattice bosons coupled to an array of atomic quantum dots}
\author{Florian Magnus Dobler}
\affiliation{Department of Physics, Royal Holloway, University of London,
Egham, Surrey TW20 0EX, United Kingdom}
\affiliation{Fachbereich Physik, Universit\"at Konstanz, D-78457 Konstanz, Germany}
\author{Anna Posazhennikova}
\affiliation{Department of Physics, Royal Holloway, University of London,
 Egham, Surrey TW20 0EX, United Kingdom}


\date{\today}

\begin{abstract}

We analyze quantum phase transitions in a system of optical lattice bosons coupled to an array of atomic quantum dots. Atomic quantum dots are represented by hard-core bosons of different hyperfine species and therefore can be mapped onto pseudospins-1/2.
The system parallels the Bose-Hubbard model with an additional assisted tunneling via coupling to the atomic quantum dots. We calculate the phase diagram of the combined system, numerically within the Gutzwiller ansatz and analytically using the mean-field decoupling approximation. The result of the assisted Bose-Hubbard model is that the Mott-superfluid transition still takes place, however, the Mott lobes strongly depend on the system parameters such as the detuning. One can even reverse the usual hierarchy of the lobes with the first lobe becoming the smallest. The phase transition in the bosonic subsystem is accompanied by a magnetization rotation in the pseudospin subsystem with the tilting angle being an effective order parameter. When direct tunneling is taken into account, the Mott lobes can be made disappear and the bosonic subsystem becomes superfluid throughout. 

\end{abstract}



\maketitle

\section{Introduction}

The celebrated Bose-Hubbard model (BHM) describes a system of interacting bosons with hopping between nearest neighbor sites \cite{Fisher1989}. This model exhibits a quantum phase transition from the incompressible Mott insulator state to the compressible superfluid state as a result of the competition between the kinetic energy and the on-site repulsion. Once it was understood that such model could be realized in optical lattices of cold bosons in a controlled way \cite{Jaksch1998} it did not take much time to experimentally observe the Mott-superfluid quantum phase transition \cite{Greiner2002,Greiner2002_2}.

After the experimental realization in optical lattices, the model became extremely popular, having been explored and studied under various conditions and modifications. The topic of non-standard Bose-Hubbard models became a separate research branch motivated by further experimental advances \cite{Dutta2015}. Various additional terms (e.g. density-induced tunneling, three-body interactions, dipolar interactions, interactions with fermions) and features of bosons (e.g. spin) were taken into account. Unsurprisingly, these resulted in modifications of the typical Bose-Hubbard model phase diagram, with new phases appearing and Mott lobes being modified in size and shape \cite{Dutta2015}.
 
In this work we study another non-standard Bose-Hubbard model where the usual, direct tunneling between the sites is suppressed and instead an assisted tunneling due to the coupling to atomic quantum dots (AQDs), which are described as pseudospin-1/2, is taken into account.  Assisted tunneling, hopping, coupling via different excitations or particles are not rare in solid state physics and condensed matter with many examples ranging from, for instance, phonon-assisted tunneling in semiconductors \cite{Kleinman1965} and electron chains \cite{Emin1987} to the celebrated Anderson model where conduction electrons of a metal hybridize with d-electrons of magnetic impurities \cite{Anderson1961,Hewson}.

\begin{figure}[!bt]
\begin{center}
\includegraphics[width=0.6\textwidth]{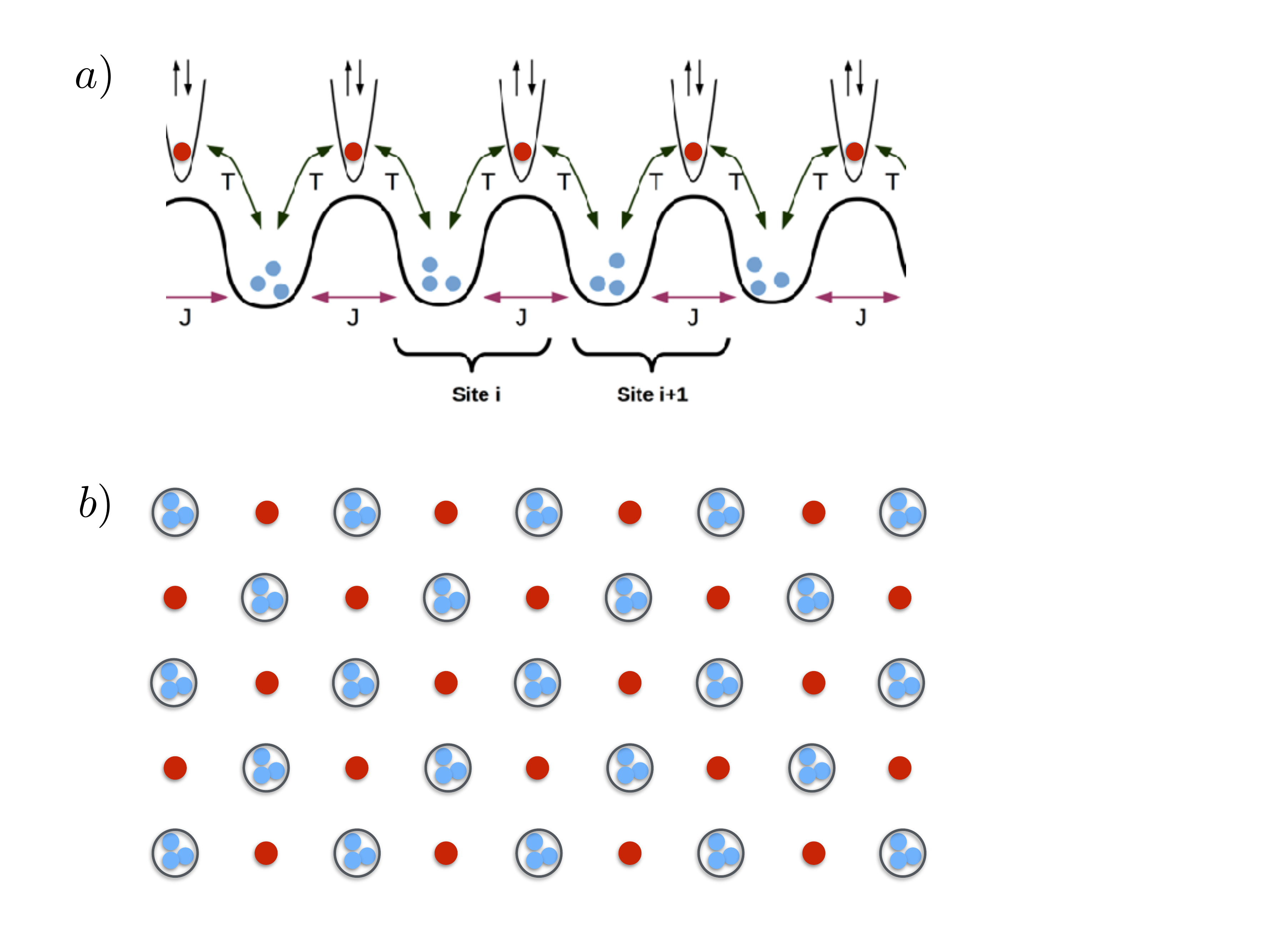}\vspace*{-0.8em} 
\end{center}
\caption{  ({\it a}) One- and  ({\it b}) two-dimensional schematic setup of the system: an optical lattice of spinless bosons (blue/light-grey circles) is coupled to an array of atomic quantum dots represented by bosons of a different atomic species (red/dark-gray circles).  $J$ is hopping between the spinless boson optical lattice sites, $T$ is the coupling of the lattice bosons to AQDs due to Raman transitions (see text). }
\label{setup}
\end{figure}

In our setup AQDs are represented by hard-core bosons coupled to repulsive bosons of different atomic species in the optical lattice by laser-induced Raman transitions (Fig. \ref{setup}). Two AQDs coupled via one bosonic reservoir were considered previously and it was shown how entanglement could be generated between two AQDs, or two pseudospins due to the induced effective interaction \cite{Posazhennikova2013,Posazhennikova2009}. 
On the other hand, an AQD can induce Josephson tunneling between two condensates coupled only through the artificial impurity \cite{Fischer2008}. A quantum order-disorder phase transition was shown to take place in a system of independent AQDs coupled to a single Bose-Einstein condensate \cite{Orth2008}. Raman-assisted hopping was also used to realize the so-called anyon Hubbard model with a rich ground state physics \cite{Santos2015}. The studies point towards a possibility of quantum phase transitions within our coupled system. 

We refer to our model as assisted Bose-Hubbard model (ABHM) and treat it within the Gutzwiller ansatz \cite{Gutzorigin1965}, which allows calculation of the superfluid order parameter and pseudospin components, depending on chosen chemical potential and interaction. We also perform an analytical calculation of the phase boundary based on the mean-field decoupling approximation \cite{Oosten2001}. Both approaches are in perfect agreement with each other. We find that the Mott-superfluid transition does take place in the bosonic subsystem. However, the Mott-superfluid phase boundary strongly depends on the system parameters to the extent that the usual Mott lobe hierarchy with the first lobe being the largest can be reversed. If direct hopping is included, we show it contributes to the reduction of the lobes with their disappearance for large enough values of the direct hopping.  The Mott - superfluid transition is accompanied by the pseudospin rotation from $z$- to $x$-axis, so that the tilting angle with $z$-axis becomes finite, once the bosonic subsystem is superfluid. We study in detail how the phase transitions can be modified and manipulated in our model. Although mean-field approaches are known to be exact only when the spatial dimension is infinite, they provide a good qualitative description of the quantum phase transition and are well suited for our purposes of describing drastic changes due to parameter manipulations. For exact phase diagram of the system, one should invoke methods such as Quantum Monte Carlo, for example \cite{Svistunov2007}. 


\section{The model}

\begin{figure}[!bt]
\begin{center}
\includegraphics[width=0.5\textwidth]{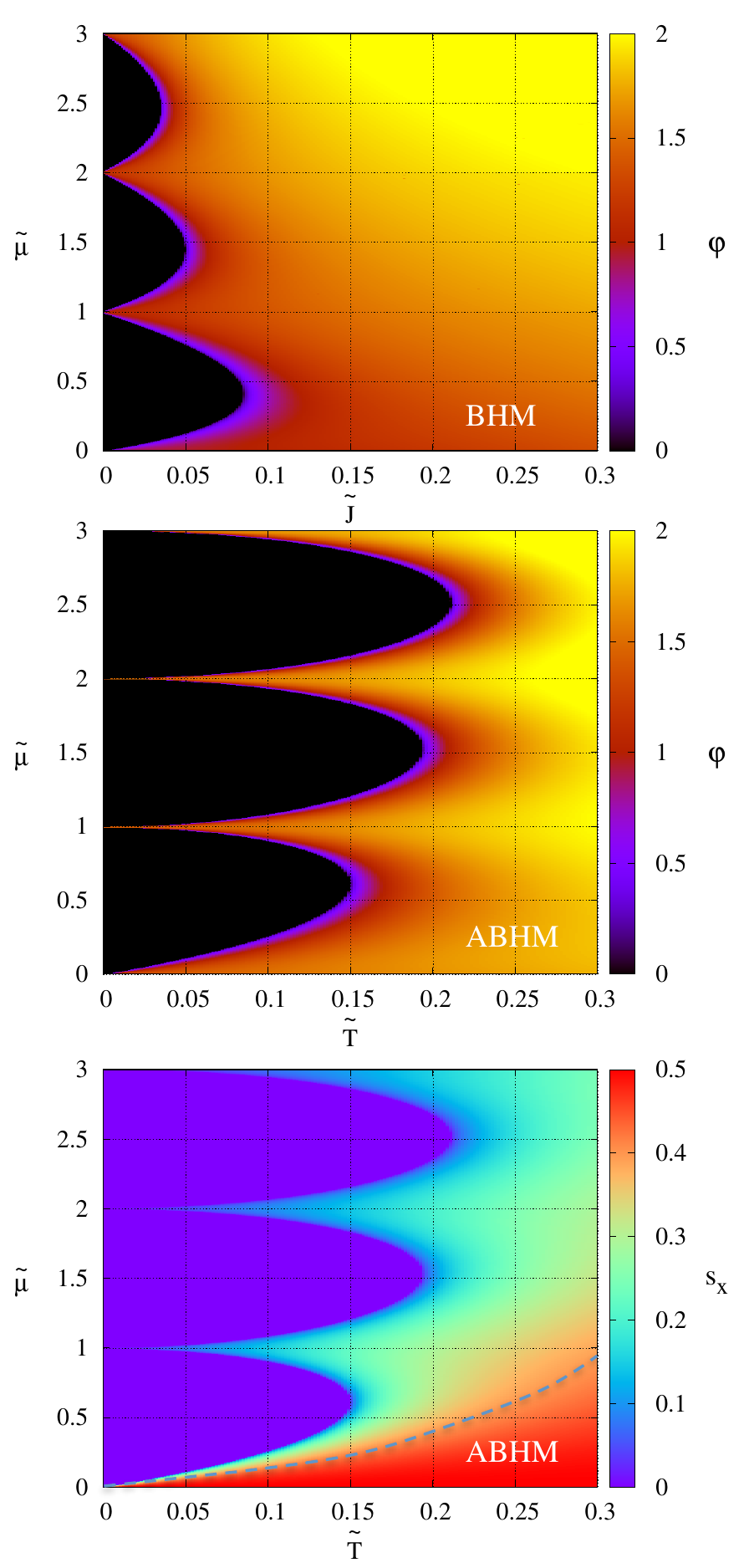} 
\end{center}
\caption{Phase diagram for the BHM (T=0), and assisted Bose Hubbard model ($J=0$ and detuning $\delta=0$), described by Hamiltonian \eqref{Ham}. The first two panels show the superfluid order parameter $\varphi$ while the lowest panel shows the $x$-component of the pseudospin $s_x$. All energies are expressed in units of $U$:  $\tilde \mu=\frac{\mu}{U}$, $\tilde T=\frac{T}{U}$, and $\tilde J=\frac{J}{U}$. The dashed line is the visual guide marking the area on the diagram where $s_x$ has a value of $1/2$ or is very close to it.}
\label{comparison}
\end{figure} 

We consider spinless bosons coupled to an array of atomic quantum dots as shown in Fig. \ref{setup}. Fig. \ref{setup} exemplifies a one-dimensional system, whereas the extension to higher dimensions is trivial. 
The general Hamiltonian of the system  is the following
\bea
H=&-&J\sum_{\langle i,j \rangle}\hat a_i^{\dagger}\hat a_j+\frac{U}{2}\sum_i\hat n_i(\hat n_i-1)-\mu \sum_i\hat n_i \nn \\
&+&T\left[\sum_i(\hat a_i^{\dagger}\hat \sigma_-^i+h.c.)+\sum_i(\hat a_{i+1}^{\dagger}\hat \sigma_-^{i}+h.c.)\right] \nn \\ &-&(\delta +\mu)\sum_i\frac{1+\hat \sigma_z^i}{2}.
\label{Ham}
\ea
The first three terms constitute the standard Bose-Hubbard model. Here  $\hat a_i^{\dagger}, \hat a_i $ are creation and annihilation operators of bosons on site $i$, $\hat n_i=\hat a^{\dagger}_i\hat a_i$, $J$ is the direct hopping between nearest neighbors, $U$ is repulsive interaction and $\mu$ is chemical potential. 

The terms proportional to $T$ describe the laser-induced hopping between the lattice sites and the atomic quantum dots. The quantum dots are described by operators $\hat b^{\dagger},\hat b$ which represent bosons of another hyperfine species than the bosons in the optical lattice, trapped in a narrow potential with an effective on-site repulsion $U_b\rightarrow \infty$. Laser-induced Raman transitions can couple the atomic quantum dots to the lattice bosons. $T$ is then proportional to the Rabi frequency $\Omega_R$ of the Raman transition between the internal atomic states of atoms in the lattice and the AQD
\beq
T=\hbar \Omega_R \int d^3r\psi_i({\bf r})\psi_b^k({\bf r}), \quad k\in i,i-1.
\eq
This kind of coupling was first introduced in Ref. \cite{Recati}. 

 In the limit of $U_b\rightarrow \infty$ we can map the bosonic operators on the dots onto pseudospin-1/2 operators  $\hat b\rightarrow \hat \sigma_-$, $\hat b^{\dagger}\rightarrow \hat \sigma_+$ and $\hat b^{\dagger}\hat b=(1+\hat \sigma_z)/2$ as it was originally done in Ref. \cite{Matsubara1956}. Here $ \hat \sigma_+=(\hat \sigma_x+i\hat \sigma_y)/2$,   $ \hat \sigma_-=(\hat \sigma_x-i\hat \sigma_y)/2$, where  $\hat \sigma_x, \hat\sigma_y, \hat\sigma_z$ are Pauli matrices. 
The wave function $\psi_i({\bf r})$ describes bosons on the $i$-site of the optical lattice, and $\psi_b^k$ is the wave function of the boson in the atomic quantum dot. We assume that the quantum dot wave-function overlaps only with wave-functions $\psi_i({\bf r})$ of the nearest neighbor sites. In this way, there is no direct interaction between the pseudospin sites, but only an induced one. The parameter $\delta$ describes the detuning necessary to suppress spontaneous emission at the AQD sites. 

\subsection{Gutzwiller ansatz approach}

We now aim to describe the quantum phase transitions in the ABHM for various detunings using the Gutzwiller ansatz.
The version of Gutzwiller ansatz for a system of bosons in a lattice \cite{GutzBosons1991} is a product of Fock states
\beq
\label{BHM_ansatz}
\ket{\Psi}_{BH} = \prod_{j}^{N}\sum_{n=0}^{\infty} f^{(j)}_{n} \ket{n}_j ,
\eq
where $N$ is the total number of lattice sites and $f^{(j)}_{n}$ are coefficients of a site $j$ with a $n$ number of bosons. Since just two states describe the state of a quantum dot, we can assign spin-up (occupied dot) and spin-down (empty dot) notations for them, so that one AQD is described by the state 
\beq
a_1\ket{\uparrow} +a_0\ket{\downarrow}.
\label{dot}
\eq
The state of the coupled system is then
\beq
|\Psi \rangle=\prod_{i}^{N}\left[\sum_{n=0}^{\infty}f_n^{(i)}\ket{n}_i\otimes (a_1^{(i)}\ket{\uparrow}_i +a_0^{(i)}\ket{\downarrow}_i) \right],
\eq
where $i$ numbers the lattice sites (see Fig. \ref{setup}). 
All the coefficients are normalized:
\beq
\sum_n|f_n^{(i)}|^2=1, \quad |a_0^{(i)}|^2+|a_1^{(i)}|^2=1. 
\eq
We assume that the coefficients $f_n$ are real, however, we need $a_0$ and $a_1$ to be complex. We can now approximate the lattice sites with as many as seven states, depending on the chemical potential.  Evaluating the Hamiltonian locally we get 
\bea
\langle  \Psi | H | \Psi \rangle&=&-zJ\left[\sum_{n=0}^{n_0+2}\sqrt{n+1}f_{n}f_{n+1}\right]^2 \nn \\
&+&\frac{U}{2}\left[\sum_{n=0}^{n_0+3}f_{n}^2(n^2-n)\right] \nonumber \\
&+&zT [a_1^*a_0+a_1a_0^*]\left[\sum_{n=0}^{n_0+2}\sqrt{n+1}f_{n}f_{n+1}\right] \nn \\
&-&\mu\left[ |a_1|^2 + \sum_{n=0}^{n_0+3}nf_n^2\right]-\delta |a_1|^2.
\ea
Here we introduced $z=2d$ the number of nearest neighbors in terms of the spatial dimension $d$, 
$n_0$ is the average occupation of optical lattice site per lobe. 
In this way we got the local energy in terms of the coefficients of the Gutzwiller ansatz. In order to reduce the number of parameters, we divide everything through $U$ and introduce $\tilde \mu=\mu/U$, $\tilde \delta =\delta/U$, $\tilde T=T/U$ and $\tilde J=J/U$. The local order parameters can be expressed in terms of the coefficients as follows
\bea
\varphi &=&\langle \hat a_i \rangle=\sum_{n=0}^{n_0+2}\sqrt{n+1}f_nf_{n+1} , \nn \\
s_x&=&\frac{1}{2}\langle \hat \sigma_x^i \rangle=\frac{1}{2}(a_0^*a_1+a_1^*a_0), \nn \\
s_y&=&\frac{1}{2}\langle \hat \sigma_y^i \rangle=\frac{i}{2}(a_0^*a_1-a_1^*a_0), \nn \\
s_z&=&\frac{1}{2}\langle \hat \sigma_z^i \rangle=\frac{1}{2}(|a_1|^2-|a_0|^2).
\ea

\begin{figure}[!bt]
\begin{center}
\includegraphics[width=0.5\textwidth]{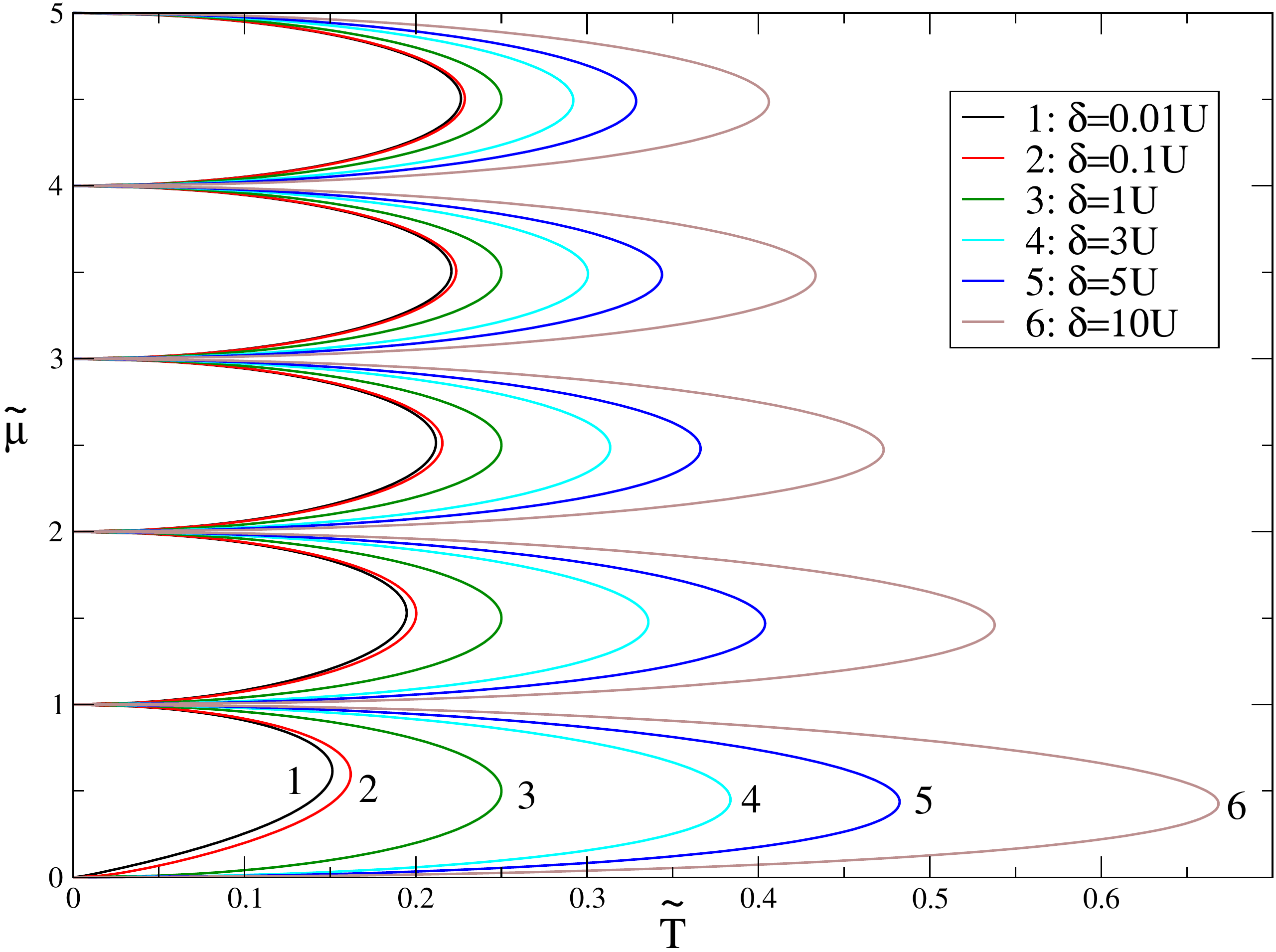} 
\end{center}
\caption{First five lobes of the Bose-Hubbard model phase boundary with assisted hopping only (direct hopping $J=0$) for different detuning $\delta$. Inside the lobes the system is in the Mott state ($\varphi=0$ and $s_x=0$), outside the lobes the system is superfluid so that  $\varphi\neq 0$ and $s_x\neq 0$.  }
\label{5_lobes}
\end{figure} 

 The phase diagram will be obtained by numerical minimization of the local energy w.r.t. the normalized coefficients for fixed values of $\tilde \delta $, $\tilde \mu$, $\tilde T$ and $\tilde J$, from which the order parameters will follow. $s_y$ turns out to always be zero as a result of these calculations and is therefore not shown anywhere. 

\subsection{Mean field description}

We can also calculate the phase boundary using a simple decoupling approximation \cite{Oosten2001}. Namely, we assume the mean-field decoupling in the terms proportional to $T$ and $J$
\bea
\hat a_i^{\dagger}\hat \sigma_-^i&\approx& s_x \hat a_i^{\dagger}+\varphi \hat \sigma_-^i -\varphi s_x, \nn \\
\hat a_i^{\dagger}\hat a_j &\approx& \varphi ( \hat  a_i^{\dagger}+  \hat a_j ) -\varphi^2,
\ea
which amounts to first order contributions in fluctuations in the order parameters. 
This kind of decoupling allows us to deal with the effective single site Hamiltonian
\bea
H_i=&\frac{U}{2}\hat n_i(\hat n_i-1)-\mu \hat n_i -(\delta +\mu)\frac{1+\hat \sigma_z^i}{2} +zJ\varphi^2 \nn \\
+&(zTs_x-zJ\varphi) (\hat a_i^{\dagger}+\hat a_i) +zT\varphi \hat \sigma_x^{i}-2zT\varphi s_x . 
\ea

Treating the single site Hamiltonian with second-order perturbation theory, we expand the ground state energy per site to second order in $\varphi$ and $s_x$ (in the following we will omit the site index $i$).

Minimising the unperturbed single site Hamiltonian gives the ground state energy as
\beq
E^{(0)}_{n}=
\begin{cases}
    \frac{U}{2} n( n-1)-\mu n-(\mu +\delta) \quad  \text{if } \mu+\delta \geq 0, \\
    \frac{U}{2} n( n-1)-\mu n  \qquad \qquad \quad \  \text{if} \,\, \mu+\delta \leq0.
\end{cases}
\label{GS}
\eq
The chemical potential should also satisfy the inequality $U(n-1)<\mu<Un$. For $\mu+\delta>0$ all AQDs are occupied, whereas for $ \mu+\delta<0$ all AQDs are empty in the ground state.
For $\mu+\delta=0$ the two energies coincide as the AQD is in the superposition state of being occupied and empty, we refer to this condition as the "degeneracy line" in our plots. 

Each ground state energy generates an equation for the phase boundary, above and below the degeneracy line. Both cases are accounted for with the absolute value of $|\mu+\delta|$.

First order corrections to the energy are zero, as expected, whereas the second-order correction contains the following contributions
\bea
E^{(2)}_{n}=z^2T^2Fs_x^2-2zT(1+zJF) \varphi s_x \nn \\
+z\left(J(1+zJ F)-\frac{zT^2}{|\mu+\delta|}\right)\varphi^2 ,
\label{second_corr}
\ea
where
\beq
F=\frac{n+1}{\mu-Un}+\frac{n}{U(n-1)-\mu}.
\eq

The first derivative test of \eqref{second_corr} gives the critical point being $(\varphi,s_x)=(0,0)$ as expected. The Hessian matrix of second derivatives reads
\beq
\begin{pmatrix} 2z^2T^2F &  -2zT(1+zJF) \\ -2zT(1+zJF) & 2z(J(1+zJF)-\frac{zT^2}{|\mu+\delta|}) \end{pmatrix}.
\eq

The phase boundary is defined by the condition requiring the determinant of the Hessian matrix to be zero for a zero eigenvalue since this is where the determinant changes sign. For $T=0$ the boundary condition for the BHM follows

\beq
\tilde J_{BH}=-\frac{1}{z\tilde F}=\frac{1}{z}\frac{(\tilde \mu-n)(n-1-\tilde \mu)}{(\tilde \mu+1)}.
\label{cond_J}
\eq
where $\tilde F = UF$. For $J=0$ we get the condition for the ABHM
\beq
\tilde T_{ABH}=\frac{1}{z}\sqrt{-\frac{|\tilde \mu+\tilde \delta|}{\tilde F}}=\frac{1}{z}\left(\frac{|\tilde \mu+\tilde \delta|(\tilde \mu-n)(n-1-\tilde \mu)}{\tilde\mu+1}  \right)^\frac{1}{2}. 
\label{cond_T}
\eq

The general phase boundary condition can be written as
\beq
\tilde T =\frac{1}{z}\sqrt{- \frac{|\tilde \mu+\tilde \delta|(1+z\tilde F \tilde J)}{\tilde F}}, 
\label{gen_cond}
\eq
or, equivalently, as
\beq
\tilde J =-\frac{1}{z\tilde F} - \frac{z \tilde T^{2}}{|\tilde \mu+\tilde \delta|}
\label{gen_cond__2}
\eq
which reduces to \eqref{cond_T} for $J=0$ and to \eqref{cond_J} for $T=0$. In the limit of large $\tilde \mu$, $\tilde J_{BH}\rightarrow 0$ while $\tilde T_{ABH}\rightarrow ((\tilde \mu-n)(n-1-\tilde \mu)  )^{1/2} /z$ identical to the case of $\tilde \delta =1$ shown in Fig. \ref{5_lobes}. The square root is a second order polynomial with periodic boundary conditions and a maximum value of $1/2$. For the one-dimensional case plotted, the Mott lobes tend towards becoming symmetric with a maximum value of $\tilde T_{c}=1/4$ in the large $\tilde \mu$ limit (also need $\tilde \mu + \tilde \delta >0$). Furthermore one should take into account that for large occupancies the Bose-Einstein condensation will take place, hence the numerical and analytical results are not valid in this regime.


\section{Results}

\subsection{Phase diagrams for different detuning $\delta$}

\begin{figure}[!bt]
\begin{center}
\includegraphics[width=0.5\textwidth]{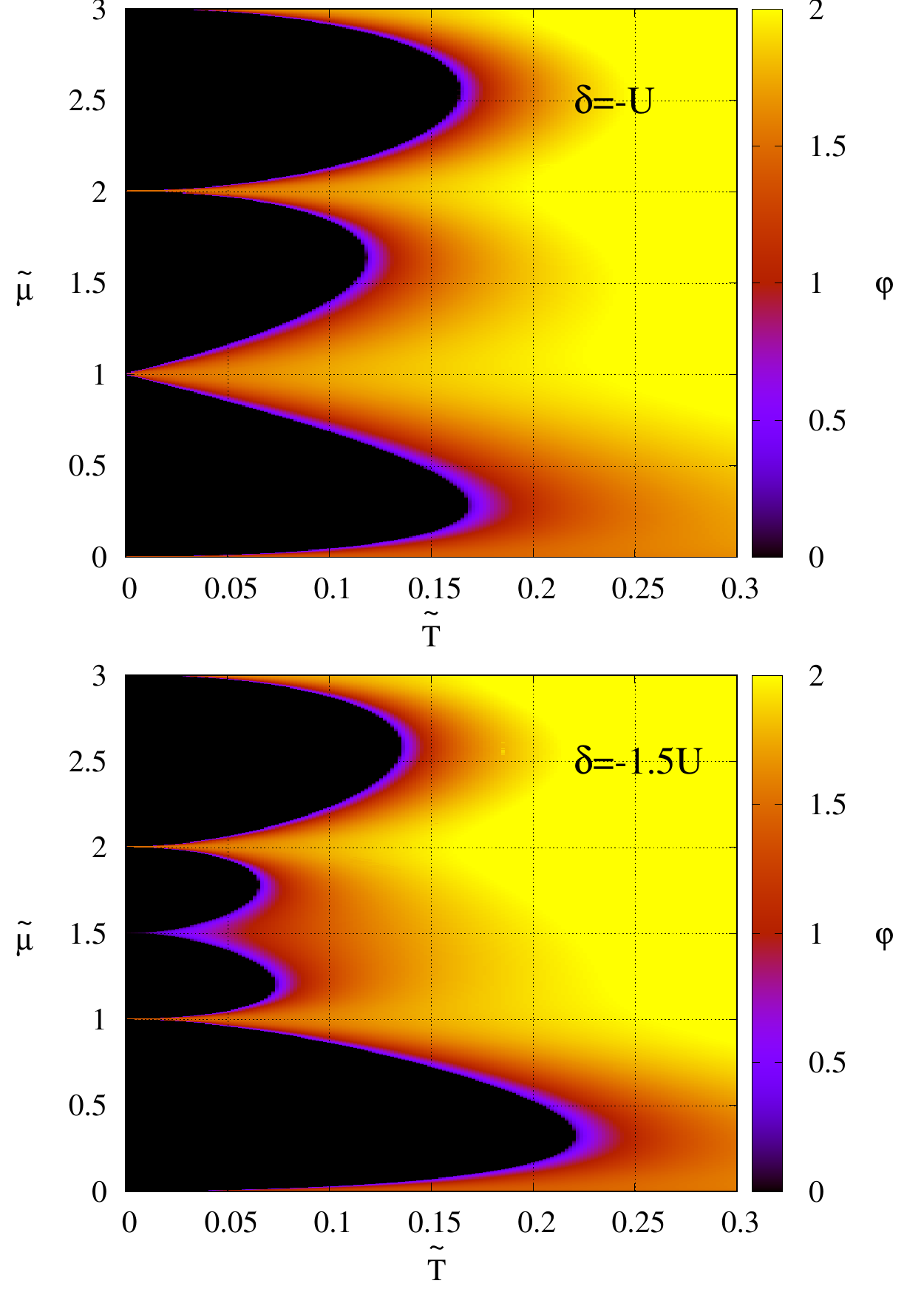} 
\end{center}
\caption{Phase diagram of the ABHM for negative detuning ($\tilde \delta =-1$ and $\tilde \delta=-1.5$). Inside the lobes the superfluid order parameter $\varphi=0$, whereas outside the lobes $\varphi$ is finite.}
\label{neg_delta}
\end{figure} 

\begin{figure}[!bt]
\begin{center}
\includegraphics[width=0.5\textwidth]{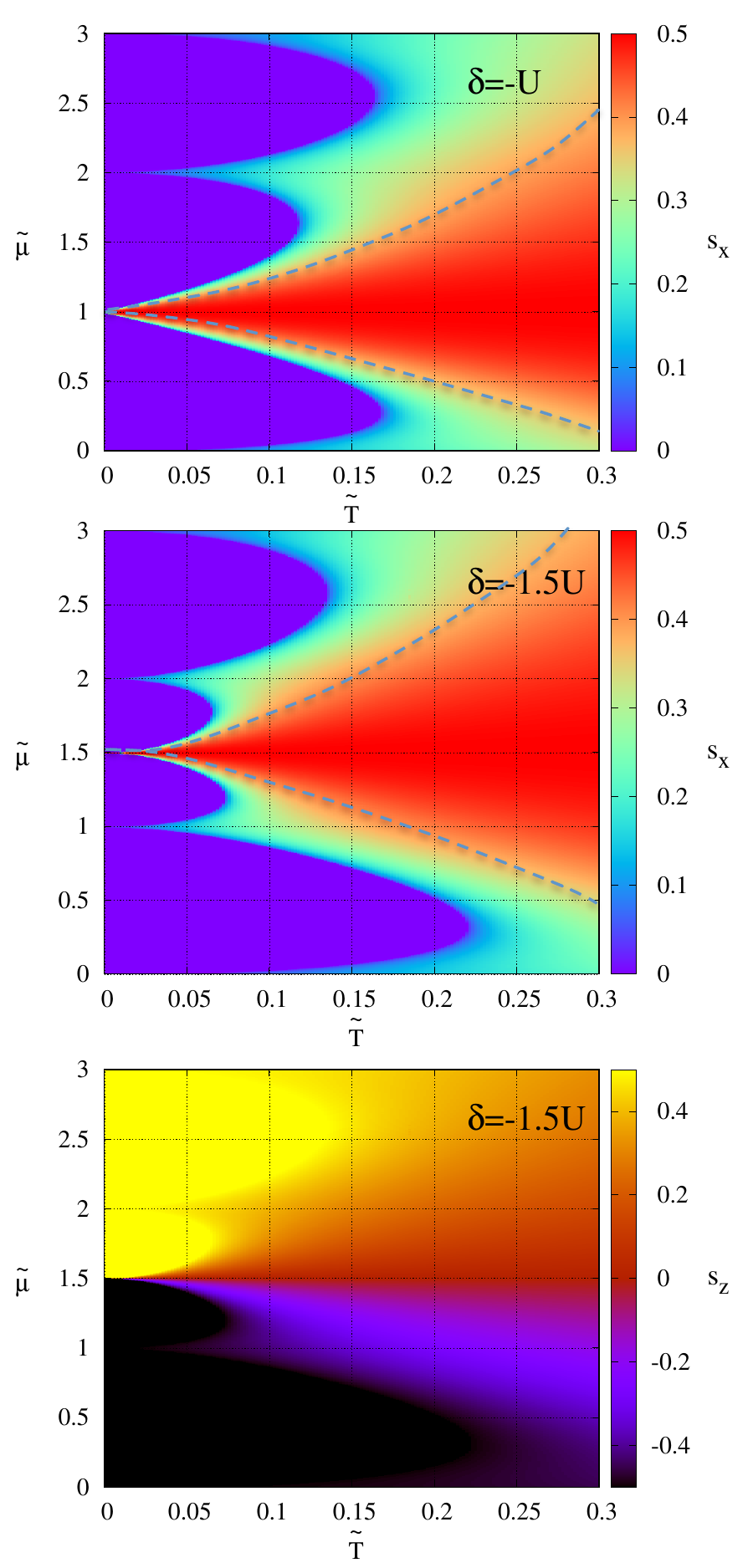} 
\end{center}
\caption{Phase diagram of the AQD subsystem for negative detuning $\tilde \delta=-1$ and $\tilde \delta=-1.5$. The size of  $s_x$ is shown for $\tilde \mu=\mu/U$ versus $\tilde T=T/U$.  Violet indicates zero $s_x$ inside the Mott lobes. $s_x$ is maximum  along the  energy degeneracy line $\mu+\delta=0$. The lowest panel shows $s_z$ which changes sign as it crosses the "degeneracy line". $s_z$ has its maximum value $|s_z|=0.5$ inside the Mott lobes. The dashed line is the visual guide marking the area on the diagram where $s_x=1/2$ or is very close to it.}
\label{spins_neg}
\end{figure} 

\begin{figure}[!bt]
\begin{center}
\includegraphics[width=0.4\textwidth]{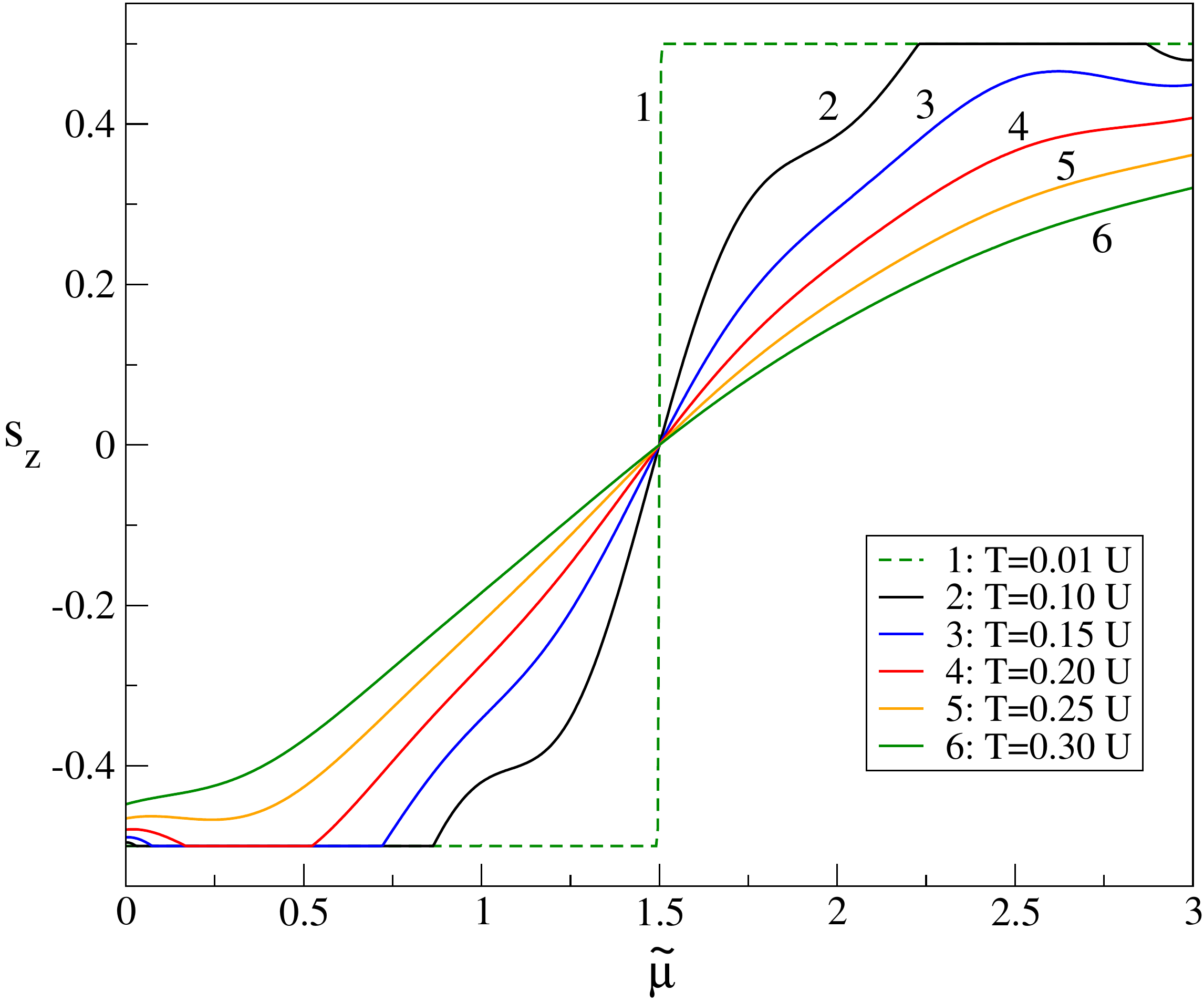} 
\end{center}
\caption{The z-component of the pseudospin $s_z$ versus $\tilde \mu=\mu/U$ for various values of $T$ and fixed $\tilde \delta=-1.5$. In terms of occupation numbers of AQDs, $s_z=-0.5$ corresponds to an unoccupied dot, whereas $s_z=0.5$ to singly-occupied AQD.   }
\label{spin_z}
\end{figure}

All results presented in the section are for $z=2$, i.e. for one-dimensional case. For higher dimensions the results are trivially rescalable (see Table \ref{tips}). 
We start with comparing the standard phase diagram for BHM with the phase diagram for our model with $J=0$ and $\delta=0$, so that only assisted hopping is present in the system. The results are shown in Fig. \ref{comparison}.  We notice several interesting features. First of all, the Mott-superfluid transition does take place in our model with the Mott lobes being in general larger than the BHM ones, which is expected since the assisted hopping is effectively weaker than direct hopping $J$. Secondly,  the usual lobe hierarchy, at least for the first several lobes, is reversed, with the first lobe being the smallest. Thirdly in the large $\mu$ limit, the lobes tend towards a fixed value of $T_c=1/4$ while in BHM due to the costly onsite interaction for large densities $J_c=0$. Lobes are broader in our case and the first lobe is slanted upwards. With increasing dimension, the lobes decrease but remain larger than those in BHM (see Table \ref{tips}). 

\begin {table}
\begin{tabular}{|c|c|c|c|}
  \hline
  Lobe & d=1 & d=2 & d=3 \\ \hline
  1.      & 0.150 (0.086) & 0.075 (0.043) & 0.050 (0.029) \\
  2.     & 0.194 (0.051) & 0.097 (0.025) & 0.065 (0.017) \\
  3.     &  0.211 (0.036) & 0.106 (0.018) & 0.070 (0.012) \\
  \hline
 \end{tabular}
 \caption{Table showing values for the tips of the first three Mott lobes in the ABHM for dimensions $d=1$, $d=2$ and $d=3$. In brackets, Mott lobe tips for Bose-Hubbard model are shown for comparison. The results are from mean-field. Parameters $J=0$ and $\delta=0$. }
 \label{tips}
 \end {table}

To understand what is going on in the pseudospin subsystem we plot $s_x$ versus $\tilde \mu$ and $\tilde T$ in Fig. \ref{comparison}, lower panel. We see that pseudospin behavior is correlated with the superfluid order parameter. Inside the Mott lobes $s_x=0$ and $s_z=1/2$, whereas outside the lobes $s_z$ is always smaller than $1/2$, and $s_x$ acquires a nonzero value. It means that when superfluidity in the bosonic subsystem is established, it effectively rotates the pseudospin from being aligned along the z-axis to being aligned along the x-axis. Hence, the tilting angle of the spin with the $z$-axis plays the role of an effective order parameter in the pseudospin subsystem. 

In terms of the occupation of the AQDs, inside the Mott lobes all dots are occupied, indicated by $s_z=1/2$, while outside of the lobes the decreasing $s_z$ indicates a decrease in the occupation of AQDs. $s_z=0$, or $s_x=1/2$ means the dots are in the equally weighted superposition of empty and occupied states (this area is marked by the dashed line in Fig. 2). 

We now explore how the detuning $\delta$ influences this behavior. In Fig. \ref{5_lobes}  we show the first five lobes of the ABHM phase diagram for various values of $\delta$. For small detuning $\tilde\delta<1$ the first lobe is the smallest. For $\tilde \delta=1$ the lobes are of the same size as follows from \eqref{cond_T}.   With increasing $\delta$ the usual Mott lobe hierarchy is restored, although lobes are much larger compared to those in BHM in Fig. \ref{comparison} and moreover for large $\mu$ they do not vanish (see Table \ref{tips2} for higher dimensions).

\begin {table}
\begin{tabular}{|c|c|c|c|}
  \hline
  Lobe & d=1 & d=2 & d=3 \\ \hline
  1.      & 0.669  & 0.334 & 0.223 \\
  2.     & 0.538 & 0.269 & 0.179 \\
  3.     &  0.473 & 0.237 & 0.158 \\
  \hline
 \end{tabular}
 \caption{Table showing values for the tips of the first three Mott lobes in the ABHM for dimensions $d=1$, $d=2$ and $d=3$. Parameters $J=0$ and $\delta=10$. }
 \label{tips2}
 \end {table}

When detuning is negative, the "degeneracy line" $\mu+\delta=0$ is shifted upwards to positive $\tilde \mu$-s revealing the following effect: if the degeneracy falls in between the lobes, it will "push" them apart resulting in their slanting (see Fig. \ref{neg_delta}, upper panel). When the degeneracy is inside a lobe, it will split it into two smaller lobes (see Fig. \ref{neg_delta}, lower panel) each lobe has a different AQD state due to different ground state energy.

In the pseudospin subsystem $s_x$ component of the pseudospin will have a maximum value along the degeneracy line, because $s_z=0$ for $\mu+\delta=0$ (see also Fig. \ref{spin_z}). This explains the red (gray inside dashed lines) "cones" in Fig. \ref{spins_neg} and a part of such a cone in the phase diagram for $s_x$ in Fig. \ref{comparison} (lowest panel). The dashed lines in Fig. \ref{spins_neg} mark the areas inside which $s_x$ is $1/2$ or very close to it corresponding to the equally weighted superposition of occupied and unoccupied states (the lines are guides to the eye). 
The lowest panel of Fig. \ref{spins_neg} displays $s_z$, which changes its sign, effectively flips while crossing the "degeneracy line".  In Fig. \ref{spin_z} we reveal how it happens with increasing $\tilde T$. The flat regions in Fig. \ref{spin_z} correspond to the Mott lobes when $s_z$ is exactly $1/2$ or $-1/2$ depending on the value of chemical potential.

\subsection{The effect of direct hopping $J$}

\begin{figure}[!bt]
\begin{center}
\includegraphics[width=0.5\textwidth]{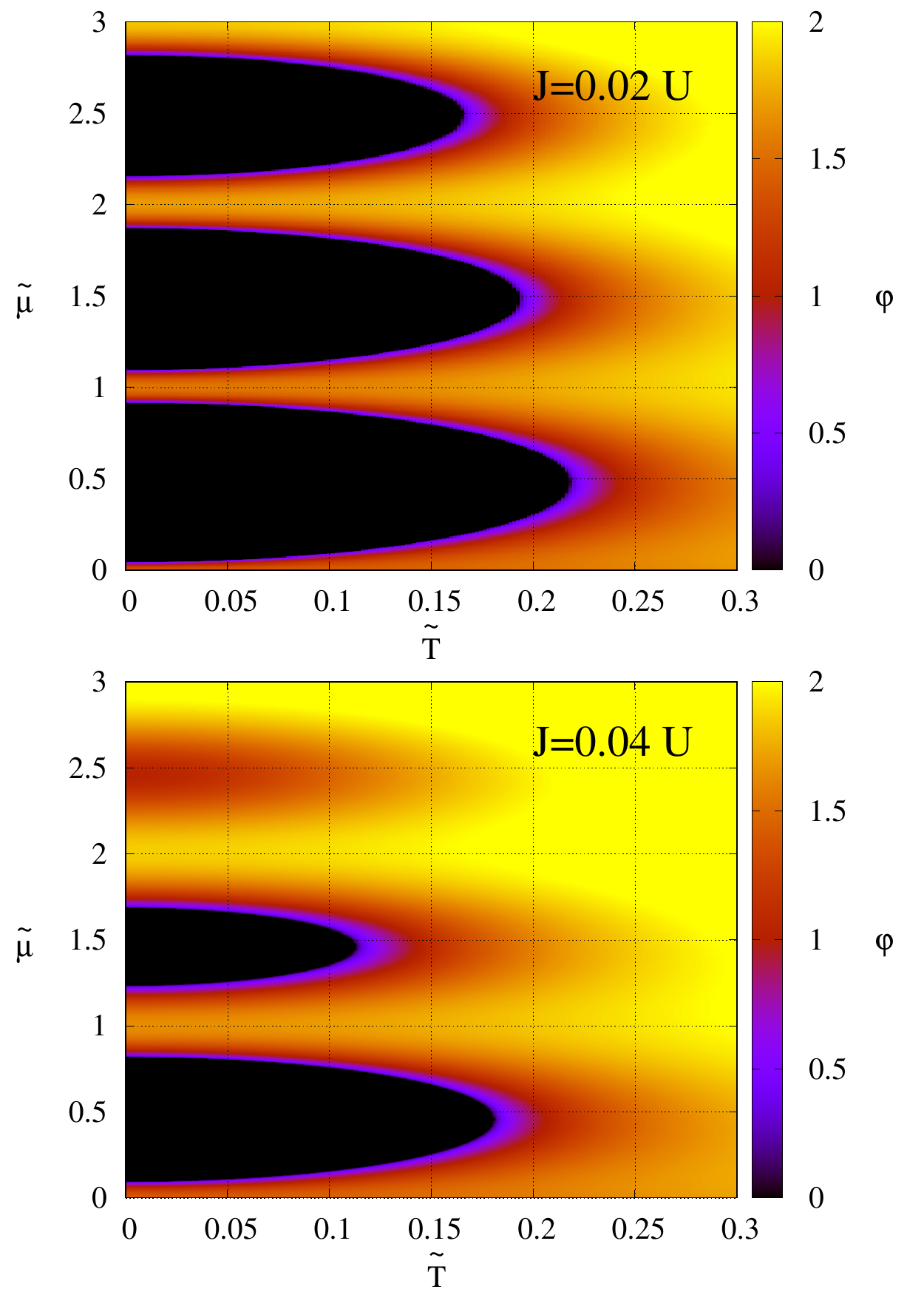} 
\end{center}
\caption{The phase diagram of the ABHM for fixed $\tilde \delta =1$ and two different $J$-s: $\tilde J=0.02$ and $\tilde J=0.04$. }
\label{effect_J}
\end{figure} 

\begin{figure}[!bt]
\begin{center}
\includegraphics[width=0.5\textwidth]{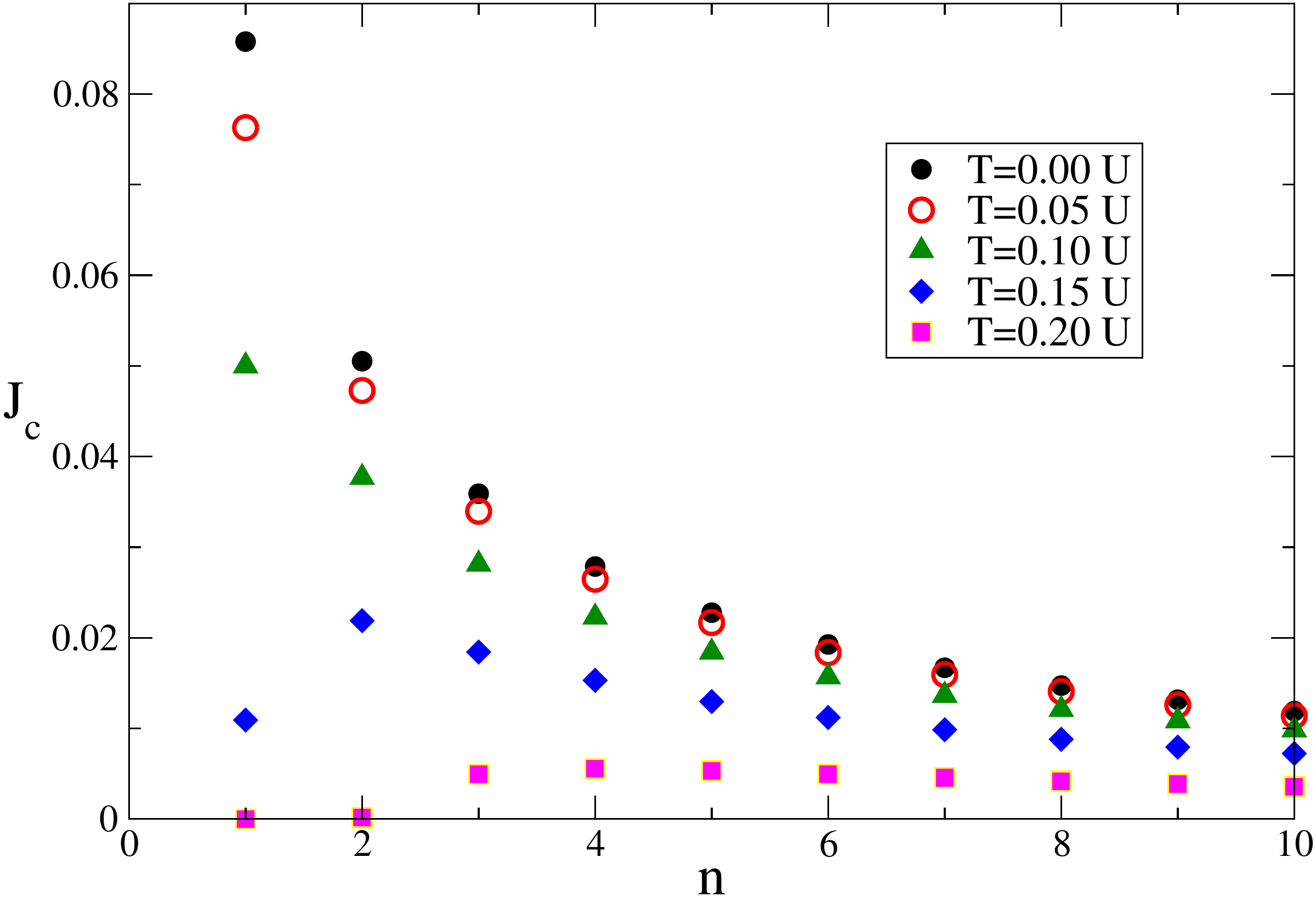} 
\end{center}
\caption{Tips of the Mott lobes, $J_c$, for different values of the assisted coupling $T$, $\tilde \delta=0.1$ for all curves. }
\label{J_crit}
\end{figure}

We now explore the effect of $J$ on the phase diagram of ABHM. One can see from Eq. \eqref{gen_cond} that $J$ suppresses $T$ and therefore reduces the Mott lobes.  It follows from Eq. \eqref{gen_cond} that the minimum value of $J$ for which mottness will be completely suppressed is equal to $J_c$, the values for the Mott lobe tips in the standard BHM. The gradual disappearance of the Mott lobes is demonstrated in Fig. \ref{effect_J} for $\tilde \delta=1$ and two different values of $\tilde J$: $0.02$ and $0.04$. The lobes become narrower and eventually disappear starting from higher lying lobes. Interestingly, when the third lobe completely disappears for $\tilde J=0.04$, there is still a "shadow" of it in the system manifested in the area of a decreased superfluid order parameter. 

The behavior of pseudospins remains similar to that in Figs. \ref{comparison} and \ref{spins_neg}, so we do not show it here.

In Fig. \ref{J_crit} we show how the tips of the Mott lobes of BHM (denoted by $J_c$) are modified by the coupling $T$ to AQDs for a fixed $\delta=0.1$. The effect of the additional coupling is most pronounced for the first lobe, one can also see how the general trend of the first lobe being the largest reverses for larger $T$. This behavior shows that direct hopping is effectively enhanced due to the assisted coupling $T$ and therefore promotes superfluidity.

\section{Conclusions}

We consider an effective hybrid system of coupled bosons and AQDs or qubits. Bosons can be coupled to each other via qubits, and vice versa. We show that in this case the bosonic subsystem still undergoes a Mott-superfluid quantum phase transition which, unlike the transition in an optical bosonic lattice alone, can be manipulated. Two main parameters in the system are the assisted coupling and the detuning.  We show that depending on their values, hierarchy and Mott lobes sizes can be strongly affected. Specifically, for finite $\delta$ the lobes become even larger, with the system favoring the mottness state. For finite but negative detuning the degeneracy line of the ground state energy is shifted upwards resulting in the splitting of the corresponding Mott lobes into two smaller ones.

If a sufficiently strong direct coupling between the bosonic sites is included, the bosonic subsystem becomes superfluid throughout. It is interesting if similar ideas could be applied to superconducting systems, although the task to study such effects becomes much more difficult. 

What about the qubit subsystem? Its behavior can be described in terms of an order parameter, in particular the x-component of the pseudospin $s_x$. Our calculation shows that $s_y=0$ is irrespective of what happens in the bosonic subsystem. $s_x$, however, is zero only inside the Mott lobes and finite outside when the bosonic subsystem is superfluid. It reaches maximum along the energy degeneracy line, whereas $s_z$ component demonstrates a spin flip while crossing the line. The finite superfluid parameter serves as an external magnetic field for the pseudospin subsystem. 

The system can be realized experimentally by superimposing two optical lattices as it is shown in Fig. \ref{setup}. The bosons in the sublattices should be of different hyperfine species and the Raman transitions between the species should be induced. Although we used mean-field approaches, exact calculations should not change the main conclusions and should confirm them qualitatively. The Mott-superfluid transitions should still exist or by appropriate tuning of the parameters the Mott state can be eliminated. All of this should take place for sufficiently homogeneous optical lattices. 

There may be some interesting applications in quantum computation. Recently hybrid systems of magnons (bosonic spin excitations of ferromagnetic crystal) and superconducting transmon qubits were engineered \cite{Tabuchi2015,Nakamura2017}. These systems are very promising for quantum computation because they can encode arbitrary qubit states in magnon nonclassical states. Extending the system to the set of arranged "pools" of magnons coupled with each other via qubits one can observe quantum and finite-temperature superfluid phase transitions which can be used in spin-based quantum information processing. 

In the future, it would be useful to properly consider spin correlations which are not possible in our single site approach. 

{\em Acknowledgements}  We acknowledge W. Belzig,  D. Kovrizhin and F. Nogueira for fruitful discussions related to the early stage of the work.

\end{document}